\documentstyle[preprint,aps,epsfig]{revtex}
\tightenlines
\begin{document}
\draft

\title{\begin{flushright}
          {\small IFT-P. 007/99 \,\, gr-qc/9901008}
       \end{flushright}
       Decay of protons and neutrons induced by acceleration}


\author{George E.A. Matsas and Daniel A.T. Vanzella}
\address{Instituto de F\'\i sica Te\'orica\\
         Universidade Estadual Paulista\\
         Rua Pamplona 145\\
         01405-900, S\~ao Paulo, SP\\
         Brazil}
\def\baselinestretch{1.5}
\maketitle


\begin{abstract}

We investigate  the  decay of accelerated protons and neutrons. 
Calculations are carried out in  the 
inertial and coaccelerated frames. Particle interpretation 
of these processes  are quite different in each frame  
but the decay rates are verified to agree  in both cases. 
For sake of simplicity  our calculations are performed in
a two-dimensional  spacetime since our conclusions
are not conceptually affected by this.

\end{abstract}

\pacs{ 04.62.+v, 12.15.Ji, 13.30.-a, 14.20.Dh } 
\newpage


\section{Introduction}
\label{Intro}

It is well known that according to the Standard Model the 
mean proper lifetime of neutrons is about $\tau_{n} = 887 s$ 
while protons are stable ($\tau_{p } > 1.6 \times 10^{25}$ years)~\cite{PDG}.  
This is only true, however, for {\em inertial} nucleons.  
There are a number of high-energy phenomena where
acceleration plays a crucial role (see Refs.~\cite{J1etal} and
\cite{J2etal}-\cite{IZ} for comprehensive discussions on electron
depolarization in storage rings and bremsstrahlung respectively).  
The influence of acceleration in particle decay was only considered  
quite recently~\cite{M}.
As it was pointed out, acceleration effects are not expected to
play a significant role
in most particle decays observed in laboratory.  Notwithstanding, 
this might not be
so under certain astrophysical and cosmological conditions.  
Muller has
estimated~\cite{M} the time decay of accelerated 
$ \mu^-$, $\pi^-$ and $p^+ $ via
the following processes:  
$$ 
(i) \;\;\mu^- \to e^- \bar\nu_e \nu_\mu\; , \;\;
(ii) \;\;\pi^- \to \mu^- \bar \nu_\mu\; , \;\; 
(iii)\;\; p^+ \to n \;e^+ \nu_e
,\;\;\; 
$$ 
as described in the laboratory frame.  Here we analyze in more detail
process 
$(iii)$ and the related one 
$$ 
(iv)\;\; n \to p^+ \; e^- \bar\nu_e\; \;.
$$ 
Process $(iii)$ is probably the most interesting one in the sense that the
proton {\em must} be accelerated in order 
to give rise to a non-vanishing rate.
In the remaining cases, non-vanishing rates are obtained even when the 
decaying particles 
($\mu^-$, $\pi^-$, $n$) 
are inertial.  As a first approximation,
Muller has considered that all particles involved are scalars.  Here we shall
treat 
$e^-$, $\nu_e$ 
and the corresponding antiparticles as fermions while $p^+$
and $n$ will be represented by a classical current.  This is a suitable
approximation as far as these nucleons are energetic enough to have a well
defined trajectory.  Moreover, we will analyze $\beta$- and inverse
$\beta$-decays in the coaccelerated frame in 
addition to in the inertial frame.
This is interesting because the particle content of 
these decays will be quite
different in each one of these frames.  This is a 
consequence of the fact that the
Minkowski vacuum corresponds to a thermal state of Rindler
particles~\cite{FD}-\cite{U}.  We have chosen to 
perform the calculations in a
two-dimensional spacetime because there is no conceptual loss at all.  A
comprehensive (but restricted to the inertial frame) 
four-dimensional spectral
analysis of the inverse $\beta$-decay for accelerated protons
and a discussion of its possible importance to cosmology and
astrophysics will be presented elsewhere.

The paper is organized as follows: In Section \ref{DNC}	we introduce the 
classical current which suitably describes the decay of accelerated
nucleons. Section \ref{inertial}
is devoted to calculate the $\beta$- and inverse 
$\beta$-decay rates in the inertial frame. In Section \ref{FermionRW} we
review the quantization of the fermionic field in the coaccelerated frame.  
In Section \ref{Rindler} we compute the  $\beta$- and inverse 
$\beta$-decay rates in the accelerated frame. For this purpose we must 
take into account the  Fulling-Davies-Unruh thermal 
bath~\cite{FD}-\cite{U}.  
Finally, in Section \ref{Discussions} we discuss our results
and further perspectives. 
We will use natural  units $k_{B} = c = \hbar = 1$ throughout this 
paper unless stated otherwise.
\setcounter{equation}{0}


\section{Decaying-nucleon current}
\label{DNC}

In order to describe the uniformly accelerated nucleon,
it is convenient to introduce 
the Rindler wedge. The Rindler wedge is the portion of Minkowski spacetime 
defined by $z > |t|$ where $(t,z)$ are the usual Minkowski coordinates. 
It is convenient to cover the Rindler wedge with Rindler coordinates $(v,u)$ 
which are related with  $(t,z)$ by
\begin{equation}
t = u \sinh v \;, \;\; z = u \cosh v  ,
\label{RC}
\end{equation} 
where $0<u<+\infty$ and $-\infty<v<+\infty$.
As a result, the line element of the Rindler wedge is written as
\begin{equation}
ds^2 = u^2 dv^2 - du^2  .
\label{LE}
\end{equation}

The  worldline of a uniformly accelerated particle with proper
acceleration $a$ is given in these coordinates by $u = a^{-1} = const$. 
Particles following this worldline have proper time  $\tau = v/a$. 
Thus let us describe a uniformly accelerated  nucleon through the 
vector current 
\begin{equation}
j^\mu = q u^\mu \delta (u-a^{-1}) ,
\label{C}
\end{equation}
where $q$ is a small coupling constant and
$u^\mu$ is the nucleon's four-velocity:
$u^\mu = (a,0)$  and $u^\mu = (\sqrt{a^2 t^2 + 1} , at)$ 
in Rindler and  Minkowski coordinates respectively.
 
The current above is fine 
to describe stable accelerated nucleons but must 
be improved to allow nucleon-decay processes. For this purpose,
let us  consider the nucleon  as a two-level system \cite{U}-\cite{BD}. 
In this scenario, neutrons $|n \rangle$ and
protons $|p  \rangle$ are going to be seen as excited and unexcited 
states of the nucleon respectively, and are
assumed to be eigenstates of the nucleon Hamiltonian $\hat H$:  
\begin{equation}
\hat H |n \rangle = m_n |n \rangle \;\; ,\;\;
\hat H |p  \rangle = m_{p } |p  \rangle \;\; ,
\end{equation}
where $m_n$ and $m_{p }$ are  the neutron and proton mass 
respectively. Hence, in order to consider nucleon decay processes, 
we replace $q$ in Eq.~(\ref{C}) by the Hermitian monopole 
\begin{equation}
\hat q(\tau )\equiv e^{i\hat H \tau} \hat q_0 e^{-i\hat H \tau}\;\; .
\label{Q}
\end{equation}
Here $G_F \equiv |\langle m_p | \hat q_0 | m_n \rangle |$
will play the role of the Fermi constant in the two-dimensional 
Minkowski spacetime.
As a result, current (\ref{C}) will be replaced by 
\begin{equation}
\hat j^\mu = \hat q(\tau) u^\mu \delta (u-a^{-1}) \;\;.
\label{CI}
\end{equation}
\setcounter{equation}{0}


\section{Inertial frame calculation of the $\beta$- and 
inverse $\beta$-decay for accelerated nucleons}
\label{inertial}

Let us firstly analyze the decay of uniformly accelerated protons 
and neutrons in the inertial frame (see processes $(iii)$ and
$(iv)$ in Sec.~\ref{Intro}). We shall describe 
electrons and neutrinos as fermionic fields:  
\begin{equation}
\hat \Psi(t,z)= \sum_{\sigma = \pm } \int_{-\infty}^{+\infty} dk
\left( \hat b_{k \sigma} \psi^{(+\omega)}_{k \sigma} (t,z)
     + \hat d^\dagger_{k \sigma} \psi^{(-\omega)}_{-k -\sigma} (t,z) 
\right)\;,
\label{FF}
\end{equation}
where $ \hat b_{k \sigma} $ and $ \hat d^\dagger_{k \sigma} $ 
are annihilation and creation operators of fermions
and antifermions, respectively, with momentum $k$ and
polarization $\sigma$. In the inertial frame, frequency,
momentum and mass $m$ are related as usually: 
$\omega=\sqrt{k^2+m^2}>0$. 
$ \psi^{(+\omega)}_{k \sigma} $ and $ \psi^{(-\omega)}_{k \sigma} $
are positive and negative frequency solutions of the Dirac equation
$i\gamma^\mu \partial_\mu \psi^{(\pm \omega)}_{k \sigma} 
 - m \psi^{(\pm \omega)}_{k \sigma} =0$.
By using the $\gamma^\mu$ matrices in the 
Dirac representation (see e.g. Ref.~\cite{IZ}), we find
\begin{equation}
\psi^{(\pm \omega)}_{k +} (t,z) =
 \frac{e^{i(\mp \omega t + kz)}}{\sqrt{2\pi}}
\left(
\begin{array}{c}
\pm \sqrt{(\omega \pm m)/2\omega} \\
0\\
k/\sqrt{2\omega(\omega \pm m)}\\
0
\end{array}
\right) \;\; 
\label{NM1}
\end{equation}
and
\begin{equation}
\psi^{(\pm \omega)}_{k -} (t,z) = 
\frac{e^{i(\mp \omega t + kz)}}{\sqrt{2\pi}}
\left(
\begin{array}{c}
0\\
\pm \sqrt{(\omega \pm m)/2\omega} \\
0\\
-k/\sqrt{2\omega(\omega \pm m)}
\end{array}
\right) \;\; .
\label{NM2}
\end{equation}
In order to keep a unified procedure for inertial and
accelerated frame calculations, 
we have orthonormalized modes (\ref{NM1})-(\ref{NM2})
according to the same inner product 
definition~\cite{BD}  that will be used in Sec.~\ref{FermionRW}: 
\begin{equation}
\langle 
\psi^{(\pm \omega)}_{k \sigma} , \psi^{(\pm \omega')}_{k' \sigma'} 
\rangle 
\equiv 
\int_\Sigma d\Sigma_\mu 
\bar \psi^{(\pm \omega)}_{k \sigma} \gamma^\mu 
\psi^{(\pm \omega')}_{k' \sigma'}
=
\delta(k-k') \delta_{\sigma \sigma'} 
\delta_{\pm \omega \; \pm \omega'} \;\;,
\label{IP}
\end{equation}
where $\bar \psi \equiv \psi^\dagger \gamma^0$,
$d\Sigma_\mu \equiv n_\mu d\Sigma $ with $n^\mu$ being a 
unit vector orthogonal to $\Sigma$ and pointing to the future,
and $\Sigma$ is an arbitrary spacelike hypersurface. 
(In this section, we have chosen $t=const$ for 
the hypersurface $\Sigma$.) As a consequence  canonical 
anticommutation relations for fields and conjugate momenta lead to 
the following simple anticommutation relations for 
creation and annihilation operators:
\begin{equation}
\{\hat b_{k \sigma},\hat b^\dagger_{k' \sigma'}\}=
\{\hat d_{k \sigma},\hat d^\dagger_{k' \sigma'}\}=
\delta(k-k') \; \delta_{\sigma \sigma'} 
\label{ACR}
\end{equation}
and
\begin{equation}
\{\hat b_{k \sigma},\hat b_{k' \sigma'}\}=
\{\hat d_{k \sigma},\hat d_{k' \sigma'}\}=
\{\hat b_{k \sigma},\hat d_{k' \sigma'}\}=
\{\hat b_{k \sigma},\hat d^\dagger_{k' \sigma'}\}=
0 \;\; .
\end{equation}

Next we couple minimally electron 
$\hat \Psi_e$ and neutrino $\hat \Psi_\nu$ 
fields to the nucleon current (\ref{CI}) 
according to the Fermi action
\begin{equation}
\hat S_I = \int d^2x \sqrt{-g} \hat j_\mu 
           (\hat{\bar \Psi}_\nu \gamma^\mu \hat \Psi_e +
            \hat{\bar \Psi}_e \gamma^\mu \hat \Psi_\nu ) \; .
\label{S}
\end{equation}
Note that the first and second terms inside  the parenthesis  
at the r.h.s of Eq.~(\ref{S}) vanish
for the $\beta$-decay (process $(iv)$ in Sec.~\ref{Intro}) 
and  inverse $\beta$-decay (process $(iii)$ in Sec.~\ref{Intro})
respectively.

Let us consider firstly the inverse $\beta$-decay.
The vacuum transition amplitude is given by
\begin{equation}
{\cal A}^{p  \to n}_{(iii)} =
\; \langle  n \vert \otimes \langle e^+_{k_e \sigma_e} , 
\nu_{k_\nu \sigma_\nu} \vert \;
\hat S_I \;
\vert 0 \rangle \otimes \vert p  \rangle \; .
\label{AMP}
\end{equation}
By using current (\ref{CI}) in Eq.~(\ref{S}), and acting with  
$\hat S_I  $ on the nucleon states in Eq.~(\ref{AMP}), we obtain
\begin{equation}
{\cal A}^{p  \to n}_{(iii)} =\;
G_F \int_{-\infty}^{+\infty} dt \int_{-\infty}^{+\infty} dz \;
\frac{e^{i  \Delta m  \tau}}{\sqrt{a^2 t^2 +1}}
\; u_{\mu} \; \delta(z-\sqrt{t^2+a^{-2}})
\langle e^+_{k_e \sigma_e} , \nu_{k_\nu \sigma_\nu} \vert\;
\hat{\bar \Psi}_\nu \gamma^\mu \hat \Psi_e\;
\vert 0 \rangle\; ,
\label{AMPI}
\end{equation}
where 
$
\Delta m \equiv m_n - m_{p }
$,  
$
\tau = a^{-1} \sinh^{-1} (at)
$
is the nucleon's proper time and we recall that in Minkowski coordinates 
the four-velocity is $u^\mu = (\sqrt{a^2 t^2 + 1} , at)$
[see below Eq.~(\ref{C})].
The numerical value of the two-dimensional Fermi constant $G_F$ will
be fixed further. 
By using the fermionic field 
(\ref{FF}) in Eq.~(\ref{AMPI}) and solving the integral in
the $z$ variable, we obtain  
\begin{eqnarray}
& & {\cal A}^{p  \to n}_{(iii)} 
= 
\frac{- \; (G_F /4\pi )\;\; \delta_{\sigma_e,-\sigma_\nu}}
      { \sqrt{\omega_\nu \omega_e 
      (\omega_\nu + m_\nu) (\omega_e - m_e)}}
\int_{-\infty}^{+\infty} d\tau 
e^{i( \Delta m \tau 
        + a^{-1} (\omega_e + \omega_\nu) \sinh a\tau 
        - a^{-1} (k_e+k_\nu)             \cosh a\tau)
  }
\nonumber
\\
&\times &
\{ [(\omega_\nu + m_\nu) (\omega_e - m_e) + k_\nu k_e] \cosh a\tau
-  [(\omega_\nu + m_\nu) k_e + (\omega_e - m_e) k_\nu] \sinh a\tau
\} \; .
\nonumber
\end{eqnarray}
The  differential transition rate  
\begin{equation}
\frac{d^2 {\cal P}^{p  \to n}_{\rm in}}{dk_e \; dk_\nu} =
\sum_{\sigma_e=\pm} \sum_{\sigma_\nu=\pm} 
| {\cal A}_{(iii)}^{p \to n} |^2
\end{equation}
calculated in the inertial frame will be, thus,
\begin{eqnarray}
& &\frac{d^2 {\cal P}^{p  \to n}_{\rm in}}{dk_e \; dk_\nu}
=
\frac{G_F^2}{8 \pi^2} 
\int_{-\infty}^{+\infty} d\tau_1  
\int_{-\infty}^{+\infty} d\tau_2 \;
\exp [i  \Delta m (\tau_1-\tau_2)]  
\nonumber
\\      
& &\times            
\exp [  i (\omega_e + \omega_\nu)(\sinh a\tau_1 -\sinh a\tau_2)/a
      - i (k_e + k_\nu) (\cosh a\tau_1-\cosh a\tau_2)/a)]
\nonumber
\\      
& & \times
\left( 
     \frac{(\omega_\nu + m_\nu) (\omega_e - m_e) + k_\nu k_e}
          {\sqrt{\omega_\nu \omega_e (\omega_\nu + m_\nu) (\omega_e - m_e)} }
     \cosh a\tau_1 
    -\frac{(\omega_\nu + m_\nu) k_e + (\omega_e - m_e) k_\nu }
          {\sqrt{\omega_\nu \omega_e (\omega_\nu + m_\nu) (\omega_e - m_e)} } 
     \sinh a\tau_1
\right)
\nonumber
\\      
& & \times
\left( 
     \frac{(\omega_\nu + m_\nu) (\omega_e - m_e) + k_\nu k_e}
          {\sqrt{\omega_\nu \omega_e (\omega_\nu + m_\nu) (\omega_e - m_e)} }
     \cosh a\tau_2 
    -\frac{(\omega_\nu + m_\nu) k_e + (\omega_e - m_e) k_\nu }
          {\sqrt{\omega_\nu \omega_e (\omega_\nu + m_\nu) (\omega_e - m_e)} } 
     \sinh a\tau_2
\right) \; .
\nonumber
\end{eqnarray}
In order to decouple the integrals above, 
it is convenient to introduce first  new variables $s$ and $\xi$  
such that
$
\tau_1 \equiv s + \xi/2 \;,\;\;
\tau_2 \equiv s - \xi/2\; .
$
After this, we write 
\begin{eqnarray}
& &\frac{d^2 {\cal P}^{p  \to n}_{\rm in}}{dk_e \; dk_\nu}
=
\frac{G_F^2}{4 \pi^2 \omega_\nu \omega_e} 
\int_{-\infty}^{+\infty} d s  
\int_{-\infty}^{+\infty} d \xi \;
e^{ i (  
        \Delta m \xi + 2 a^{-1} \sinh(a \xi/2) 
        [ 
        (\omega_\nu + \omega_e) \cosh as 
        -(k_\nu + k_e) \sinh as 
        ]
       ) 
  }
\nonumber
\\      
& &\times            
[(\omega_\nu \omega_e + k_\nu k_e) \cosh 2as -
 (\omega_e k_\nu + \omega_\nu k_e) \sinh 2as -
  m_\nu m_e \cosh a\xi] \; .
\label{TS}
\end{eqnarray}
Next, by defining a new change of variables:
$$
k_{e(\nu)} \to {k'}_{e(\nu)}= 
- \omega_{e(\nu)} \sinh (as)+k_{e(\nu)} \cosh (as),
$$
we are able to perform the integral in the $s$ variable, and
the differential transition rate (\ref{TS}) can be cast 
in the form
\begin{eqnarray}
\frac{1}{T} 
\frac{d^2 {\cal P}^{p \to n}_{\rm in}}{d{k'}_e \; d{k'}_\nu} 
& = &
\frac{G_F^2}{4 \pi^2 {\omega}'_e {\omega}'_\nu} 
\int_{-\infty}^{+\infty} d\xi \;  
\exp \left[ i \Delta m \xi + i 2 a^{-1} 
           ({\omega}'_e + {\omega}'_\nu) \sinh (a\xi/2)
     \right] 
\nonumber
\\      
& \times &  
     ({\omega}'_\nu {\omega}'_e  + {k'}_\nu {k'}_e 
       - m_\nu m_e\cosh a\xi)
 \;,
\label{DP2}
\end{eqnarray}
where $ T \equiv \int_{-\infty}^{+\infty} ds $ is the total  proper 
time and 
$ 
{\omega}'_{e(\nu)}\equiv \sqrt{{k'}^2_{e(\nu)} + m^2_{e(\nu)}}
$. 

The total transition rate 
$
{\Gamma}^{p  \to n}_{\rm in}={\cal P}^{p \to n}_{\rm in}/T
$
is obtained after integrating Eq.~(\ref{DP2}) in both momentum variables. 
For this purpose it is useful to make the following change of variables:
$$
{k'}_{e(\nu)} \to {{\tilde{k}}}_{e(\nu)} \equiv {k'}_{e(\nu)}/a\;,\;\;
\xi \to \lambda \equiv e^{a \xi /2}.
$$
(Note that ${{\tilde{k}}}_{e(\nu)}$ is adimensional.)
Hence we obtain
\begin{eqnarray}
{\Gamma}^{p  \to n}_{\rm in} 
& = &
\frac{G_F^2 a}{2 \pi^2}
\int_{-\infty}^{+\infty} \frac{d {\tilde{k}}_e }{\tilde\omega_e}
\int_{-\infty}^{+\infty} \frac{d {\tilde{k}}_\nu }{\tilde\omega_\nu}
\int_{-\infty}^{+\infty} \frac{d\lambda}{\lambda^{1-i2\Delta m/a}} \;  
\exp[i (\tilde\omega_e + \tilde\omega_\nu) (\lambda-\lambda^{-1})]
\nonumber
\\      
& \times & 
\left(
\tilde\omega_\nu  \tilde\omega_e  + {\tilde{k}}_\nu {\tilde{k}}_e 
- m_\nu m_e (\lambda^2+\lambda^{-2})/(2 a^2) 
\right)
\;,
\label{RI}
\end{eqnarray}
where 
$\tilde\omega_{e(\nu)}\equiv \sqrt{{\tilde{k}}^2_{e(\nu)}+m^2_{e(\nu)}/a^2}$.

Let us assume at this point $m_\nu \to 0$. In this case, using (3.871.3-4) 
of Ref. \cite{GR}, we perform the integration in $\lambda$ and obtain
the following final expression for the proton decay rate:
\begin{equation}
{\Gamma}^{p  \to n}_{\rm in} =
\frac{4 G_F^2 a }{\pi^2 e^{\pi \Delta m/a}} 
\int_{0}^{+\infty} d {\tilde{k}}_e \int_{0}^{+\infty} d {\tilde{k}}_\nu \;
K_{i 2\Delta m /a} 
\left[ 2 \left( \sqrt{{{\tilde{k}}}_e^2+m_e^2/a^2} +
 {{\tilde{k}}}_\nu  \right) \right]
\;.
\label{RIF}
\end{equation}

Performing analogous calculation for the $\beta$-decay,
we obtain for the neutron differential and total decay 
rates  the following expressions [see Eqs.~(\ref{DP2}) and (\ref{RI})]:
\begin{eqnarray}
\frac{1}{T} 
\frac{d^2 {\cal P}^{n \to p}_{\rm in}}{d{k'}_e \; d{k'}_\nu} 
& = &
\frac{G_F^2}{4 \pi^2 {\omega}'_e {\omega}'_\nu} 
\int_{-\infty}^{+\infty} d\xi \;  
\exp \left[ - i \Delta m \xi + i 2  a^{-1} 
           ({\omega}'_e + {\omega}'_\nu) \sinh (a\xi/2)
     \right] 
\nonumber
\\      
& \times & 
     ({\omega}'_\nu {\omega}'_e  + {k'}_\nu {k'}_e 
       - m_\nu m_e\cosh a\xi)
 \;,
\label{DP3}
\end{eqnarray}
and 
\begin{eqnarray}
{\Gamma}^{n  \to p}_{\rm in} 
& = &
\frac{G_F^2 a}{2 \pi^2}
\int_{-\infty}^{+\infty} \frac{d {\tilde{k}}_e }{\tilde\omega_e}
\int_{-\infty}^{+\infty} \frac{d {\tilde{k}}_\nu }{\tilde\omega_\nu}
\int_{-\infty}^{+\infty} \frac{d\lambda}{\lambda^{1+i2\Delta m/a}} \;  
\exp[i (\tilde\omega_e + \tilde\omega_\nu) (\lambda-\lambda^{-1})]
\nonumber
\\      
& \times & 
\left(
\tilde\omega_\nu  \tilde\omega_e  + {\tilde{k}}_\nu {\tilde{k}}_e 
- m_\nu m_e (\lambda^2+\lambda^{-2})/(2 a^2) 
\right)
\; .
\end{eqnarray}
By making $m_\nu \to 0$ in the expression above, we end up with
\begin{equation}
{\Gamma}^{n \to p }_{\rm in} 
= 
\frac{4 G_F^2 a }{\pi^2 e^{-\pi \Delta m/a}} 
\int_{0}^{+\infty} d {\tilde{k}}_e \int_{0}^{+\infty} d {\tilde{k}}_\nu \;
K_{i 2\Delta m /a} 
\left[ 2 \left( \sqrt{{{\tilde{k}}}_e^2+m_e^2/a^2} + 
{{\tilde{k}}}_\nu  \right) \right]
\;.
\label{RNF}
\end{equation}

In order to determine the value of our two-dimensional
Fermi constant $G_F$ we will impose the mean proper lifetime 
$ \tau_n(a) = 1/{\Gamma}^{n \to p }_{\rm in} $ of 
an inertial neutron to be $887s$ \cite{PDG}.  
By taking $a \to 0$ and integrating both sides of 
Eq.~(\ref{DP3}) with respect to the momentum variables, we obtain 
\begin{eqnarray}
\left. {\Gamma}^{n \to p }_{\rm in} \right|_{a \to 0}
& = &
\frac{G_F^2}{4 \pi^2}
\int_{-\infty}^{+\infty} \frac{d {k'}_e}{{\omega}'_e} 
\int_{-\infty}^{+\infty} \frac{d {k'}_\nu}{{\omega}'_\nu}  
\int_{-\infty}^{+\infty} d\xi \;  
\exp \left[ i 
           ({\omega}'_e + {\omega}'_\nu) \xi
     \right] 
\nonumber
\\      
& \times & 
\exp (- i \Delta m \xi )  
     ({\omega}'_\nu {\omega}'_e  + {k'}_\nu {k'}_e 
       - m_\nu m_e)
 \; .
\label{DP4}
\end{eqnarray}
Next, by performing the integral in $\xi$, we obtain 
\begin{equation}
\left. {\Gamma}^{n \to p }_{\rm in} \right|_{a \to 0}
=
\frac{2 G_F^2}{\pi}
\int_{m_e}^{+\infty} 
\frac{d {\omega'}_e}{\sqrt{{\omega'}_e^2 -m_e^2} } 
\int_{m_\nu}^{+\infty} 
\frac{d {\omega'}_\nu}{\sqrt{{\omega'}_\nu^2-m_\nu^2} }  
({\omega}'_\nu {\omega}'_e  - m_\nu m_e) 
\delta ({\omega}'_\nu + {\omega}'_e  - \Delta m)
 \; .
\label{DP5}
\end{equation}
Now it is easy to perform the integral in ${\omega'}_\nu$:
\begin{equation}
\left. {\Gamma}^{n \to p }_{\rm in} \right|_{a \to 0}
=
\frac{2 G_F^2}{\pi}
\int_{m_e}^{\Delta m - m_\nu} 
\frac{d {\omega'}_e}{\sqrt{{\omega'}_e^2 -m_e^2} } 
\frac{{\omega}'_e(\Delta m - {\omega}'_e) - m_\nu m_e}
     {\sqrt{ (\Delta m - {\omega}'_e)^2 - m_\nu^2 }}
 \; .
\label{DP6}
\end{equation}
By integrating the right-hand side of Eq.~(\ref{DP6}) 
with $m_\nu \to 0$ and imposing 
$1/\left. {\Gamma}^{n \to p }_{\rm in} \right|_{a \to 0}$ to be
$887s$, 
we obtain $G_F = 9.918 \times 10^{-13}$. 
Note that $G_F \ll 1$ which corroborates our perturbative approach. 
Now we are able to plot the neutron mean proper lifetime
as a function of its proper acceleration $a$ (see Fig.~1). 
Note that after an oscillatory regime it
decays steadily. In Fig.2  we plot the proton mean proper lifetime. 
The necessary energy to allow protons to decay is provided
by the external accelerating agent. For accelerations
such that the Fulling-Davies-Unruh (FDU)
temperature (see discussion in Sec.~\ref{Rindler}) 
is of order of $m_n + m_e - m_p$, 
i.e. $a/2\pi \approx 1.8 MeV$, we have that 
$\tau_p \approx \tau_n $ 
(see Fig.~1 and Fig.~2).  Such accelerations are considerably high. 
Just for sake of comparison,
protons at LHC have a proper acceleration of about $ 10^{-8} MeV $. 
\setcounter{equation}{0}


\section{Fermionic field quantization in a two-dimensional Rindler wedge}
\label{FermionRW}

We shall briefly review \cite{SMG} the quantization of the 
fermionic field in the accelerated frame since this will be
crucial for our further purposes. Let us consider the 
two-dimensional Rindler wedge described by the line element (\ref{LE}).
The Dirac equation in curved spacetime is 
$ (i\gamma_R^\mu \tilde \nabla_\mu -m) \psi_{\omega \sigma}=0$,
where
$\gamma_R^\mu \equiv (e_\alpha)^\mu \gamma^\alpha$
are the Dirac matrices in curved spacetime,
$\tilde \nabla_\mu \equiv \partial_\mu +\Gamma_\mu $
and 
$ \Gamma_\mu
= 
\frac{1}{8} [\gamma^\alpha,\gamma^\beta] (e_\alpha)^\lambda 
\tilde \nabla_\mu (e_\beta)_\lambda
$
are the 
Fock-Kondratenko connections. ( $\gamma^\mu$ are
the usual flat-spacetime Dirac matrices.) In the
Rindler wedge the relevant  tetrads are
$ (e_0)^{\mu}= u^{-1} \delta^\mu_0$, 
$ (e_i)^\mu = \delta^\mu_i$. 
As a consequence, the Dirac equation takes the form
\begin{equation}
i \frac{\partial \psi_{\omega \sigma}}{\partial v}=
\left( 
\gamma^0 mu - \frac{i \alpha_3}{2} - 
i u \alpha_3 \frac{\partial}{\partial u}
\right)
\psi_{\omega \sigma}\;,
\label{DEQ}
\end{equation}
where 
$\alpha_i \equiv \gamma^0 \gamma^i$. 

We shall express the fermionic field as
\begin{equation}
\hat \Psi(v,u)= \sum_{\sigma = \pm } \int_{0}^{+\infty} d\omega
\left( \hat b_{\omega \sigma} \psi_{\omega \sigma}(v,u)
       + \hat d^\dagger_{\omega \sigma} \psi_{-\omega -\sigma}(v,u) 
\right) ,
\label{PSA}
\end{equation}
where $\psi_{\omega \sigma}=f_{\omega \sigma}(u) e^{-i \omega v/a}$
are positive ($\omega >0$)  and negative ($\omega <0$)
energy solutions with respect to the boost Killing field 
$\partial / \partial v$ with polarization $\sigma = \pm$.
From Eq.~(\ref{DEQ}) we obtain
\begin{equation}
\hat H_u f_{\omega \sigma} = \omega f_{\omega \sigma} \;,
\label{ENA}
\end{equation}
where
\begin{equation}
\hat H_u \equiv  a 
\left[ 
m u \gamma^0 - \frac{i \alpha_3}{2} - 
i u \alpha_3 \frac{\partial}{\partial u}
\right]\;.
\label{HU}
\end{equation}
By ``squaring'' Eq.~(\ref{ENA}) and defining two-component spinors
$\chi_{j}$ ($j=1,2$) through 
\begin{equation}
 f_{\omega \sigma}(u) \equiv 
\left(
\begin{array}{c}
\chi_1 (u) \\
\chi_2 (u) 
\end{array}
\right) \;\; ,
\label{(4.5)}
\end{equation}
we obtain
\begin{eqnarray}
\left( 
u \frac{d}{du} u\frac{d}{du} 
\right) \chi_1 
& = &
\left[ 
m^2 u^2 + \frac{1}{4} - \frac{\omega^2}{a^2} 
\right] \chi_1
-\frac{i \omega}{a}\sigma_3 \chi_2  \;,
\label{XIa}
\\
\left( 
u \frac{d}{du} u\frac{d}{du} 
\right)  \chi_2
& = &
\left[ 
m^2 u^2 + \frac{1}{4} - \frac{\omega^2}{a^2} 
\right] \chi_2
-\frac{i \omega}{a}\sigma_3 \chi_1  \;.
\label{XIb}
\end{eqnarray} 
Next, by introducing  $\phi^\pm \equiv \chi_1 \mp \chi_2$, 
we can define $\xi^\pm $ and $\zeta^\pm $ through 
\begin{equation}
\phi^\pm \equiv 
\left(
\begin{array}{c}
\xi^\pm (u) \\
\zeta^\pm (u) 
\end{array}
\right) \; .
\label{(4.7)}
\end{equation}
In terms of these variables Eqs.~(\ref{XIa})-(\ref{XIb}) become 
\begin{eqnarray}
\left( u\frac{d}{du}u\frac{d}{du} \right) \xi^\pm 
& = &
[m^2u^2+(i\omega/a \pm 1/2)^2] \xi^\pm \; ,
\label{(4.8)}
\\
\left( u\frac{d}{du}u\frac{d}{du} \right) \zeta^\pm 
& = &
[m^2u^2+(i\omega/a \mp 1/2)^2] \zeta^\pm \; .
\label{(4.82)}
\end{eqnarray}
The solutions of these differential equations can be written
in terms of Hankel functions 
$ H^{(j)}_{i\omega/a \pm 1/2}(imu)$, $(j=1,2)$, 
(see (8.491.6) of Ref. \cite{GR}) 
or modified Bessel functions 
$ K_{i \omega/a \pm 1/2}(mu)$,
$ I_{i \omega/a \pm 1/2}(mu)$ 
(see (8.494.1)  of Ref. \cite{GR}).
Hence, by using Eqs.~(\ref{(4.5)}) and (\ref{(4.7)}), 
and imposing that the solutions  satisfy the 
first-order Eq.~(\ref{ENA}), we obtain 
\begin{eqnarray}
f_{\omega +} (u) &=&
A_+ 
\left(
\begin{array}{c}
K_{i \omega/a + 1/2}(mu) +
i K_{i \omega/a - 1/2}(mu)\\
0\\
-K_{i \omega/a + 1/2}(mu) +
i K_{i \omega/a - 1/2}(mu)\\
0
\end{array}
\right) \; ,
\\
f_{\omega -} (u)
&=&
A_- 
\left(
\begin{array}{c}
0\\
K_{i \omega/a + 1/2}(mu) +
i K_{i \omega/a - 1/2}(mu)\\
0\\
 K_{i \omega/a + 1/2}(mu) -
i K_{i \omega/a - 1/2}(mu)
\end{array}
\right) \; .
\end{eqnarray}
Note that solutions involving $I_{i \omega/a \pm 1/2}$ turn out
to be non-normalizable and thus must be neglected.
In order to find the normalization constants 
\begin{equation}
A_+=A_-=\left[\frac{m \cosh (\pi \omega/a)}{2\pi^2 a}\right]^{1/2}\; ,
\label{CONST}
\end{equation} 
we  have used \cite{BD} (see also Eq.~(\ref{IP}))
\begin{equation}
\langle 
\psi_{\omega \sigma} , \psi_{\omega' \sigma'} 
\rangle 
\equiv 
\int_\Sigma d\Sigma_\mu 
\bar \psi_{\omega \sigma} \gamma^\mu_R 
\psi_{\omega' \sigma'}
=
\delta(\omega-\omega') \delta_{\sigma \sigma'} \;\;,
\label{IP2}
\end{equation}
where $\bar \psi \equiv \psi^\dagger \gamma^0$ 
and  $\Sigma$ is set to be $v=const$.
Thus the normal modes of the fermionic field  (\ref{PSA}) are 
\begin{equation}
\psi_{\omega +} =
\left[\frac{m \cosh (\pi \omega/a)}{2 \pi^2 a}\right]^{1/2}
\left(
\begin{array}{c}
K_{i \omega/a + 1/2}(mu) +
i K_{i \omega/a - 1/2}(mu)\\
0\\
-K_{i \omega/a + 1/2}(mu) +
i K_{i \omega/a - 1/2}(mu)\\
0
\end{array}
\right) 
e^{-i\omega v/a} \; ,
\label{RNM1}
\end{equation}
\begin{equation}
\psi_{\omega -} =
\left[\frac{m \cosh (\pi \omega/a)}{2 \pi^2 a}\right]^{1/2}
\left(
\begin{array}{c}
0\\
K_{i \omega/a + 1/2}(mu) +
i K_{i \omega/a - 1/2}(mu)\\
0\\
K_{i \omega/a + 1/2}(mu) -
i K_{i \omega/a - 1/2}(mu)
\end{array}
\right) 
e^{-i\omega v/a} .
\label{RNM2}
\end{equation}
As a consequence, canonical anticommutation relations for fields and
conjugate momenta lead annihilation and creation operators to satisfy
the following anticommutation relations
\begin{equation}
\{\hat b_{\omega \sigma},\hat b^\dagger_{\omega' \sigma'}\}=
\{\hat d_{\omega \sigma},\hat d^\dagger_{\omega' \sigma'}\}=
\delta(\omega-\omega') \; \delta_{\sigma \sigma'} 
\label{ACR2}
\end{equation}
and
\begin{equation}
\{\hat b_{\omega \sigma},\hat b_{\omega' \sigma'}\}=
\{\hat d_{\omega \sigma},\hat d_{\omega' \sigma'}\}=
\{\hat b_{\omega \sigma},\hat d_{\omega' \sigma'}\}=
\{\hat b_{\omega \sigma},\hat d^\dagger_{\omega' \sigma'}\}=
0 \;\; .
\end{equation}

\setcounter{equation}{0}


\section{Rindler frame calculation of the 
$\beta$- and inverse $\beta$-decay for accelerated
nucleons}
\label{Rindler}

Now we analyze the $\beta$- and inverse $\beta$-decay of accelerated
nucleons from the point of view of the uniformly
accelerated frame.  Mean proper lifetimes must
be the same of the ones obtained in 
Sec.~\ref{inertial} but particle interpretation changes
significantly.  This is so because uniformly accelerated particles in the
Minkowski vacuum are immersed in the  FDU thermal bath
characterized by a temperature $T=a/2 \pi $~\cite{FD}-\cite{U}.  As it will
be shown, the proton decay which is represented in the inertial frame, 
in terms of Minkowski particles, by process $(iii)$ 
will be represented in the uniformly accelerated frame, in
terms of Rindler particles,  as the combination of the 
following processes:  
$$ 
(v)\;\; p^+ \; e^- \to n \; \nu\;,\; 
(vi) \;\; p^+ \; \bar \nu \to n \; e^+ \;,\; 
(vii) \;\; p^+ \; e^- \bar \nu \to n \;. 
$$ 
Processes $(v)-(vii)$ are characterized by the
conversion of protons in neutrons due to the absorption 
of $ e^- $ and $\bar \nu$, and emission of
$e^+$ and  $\nu$  from and to the FDU thermal bath.  
Note that process
$(iii)$ is forbidden in terms of Rindler particles because
the proton is static in the Rindler frame.

Let us calculate firstly the
transition amplitude for process $(v)$:
\begin{equation}
{\cal A}^{p  \to n}_{(v)} = 
\; \langle  n \vert \otimes \langle \nu_{\omega_\nu \sigma_\nu} \vert \;
\hat S_I \;
\vert e^-_{\omega_{e^-} \sigma_{e^-}} \rangle \otimes \vert p  \rangle\;,
\label{ACA}
\end{equation}
where $\hat S_I$ is given by Eq.~(\ref{S}) with $\gamma^\mu$ replaced by 
$\gamma^\mu_R$ and our current is given by
Eq.~(\ref{CI}). 
Thus, we obtain [we recall that in Rindler coordinates $u^\mu=(a,0)$]
\begin{equation}
{\cal A}^{p  \to n}_{(v)}  = 
\frac{G_F}{a} \int_{-\infty}^{+\infty} dv \;e^{i \Delta m v/a} 
\langle \nu_{\omega_\nu \sigma_\nu} \vert
\hat\Psi^\dagger_\nu  (v,a^{-1}) \hat \Psi_e (v,a^{-1})
\vert e^-_{\omega_{e^-} \sigma_{e^-}} \rangle\; ,
\label{V}
\end{equation}
where we note that the second term in the 
parenthesis of Eq.~(\ref{S}) does not contribute.
Next, by using Eq.~(\ref{PSA}), we obtain
\begin{equation}
{\cal A}^{p  \to n}_{(v)} = \frac{G_F}{a}\delta_{\sigma_{e^-},\sigma_\nu}
\int_{-\infty}^{+\infty} dv\; e^{i \Delta m v/a}
\psi^\dagger_{\omega_\nu\sigma_\nu} (v,a^{-1}) 
\;\psi_{\omega_{e^-}\sigma_{e^-}}(v,a^{-1})\;.
\label{V1}
\end{equation}
Using now Eq.~(\ref{RNM1}) and Eq.~(\ref{RNM2}) and performing the integral, 
we obtain
\begin{eqnarray}
{\cal A}^{p  \to n}_{(v)}  & = & \frac{4G_F}{\pi a}\sqrt{m_em_\nu \cosh (\pi
\omega_{e^-}/a)\cosh (\pi \omega_\nu/a)} \nonumber \\
& \times & 
Re\left[ K_{i\omega_\nu/a-1/2}(m_\nu /a)\; K_{i\omega_{e^-}/a +1/2} (m_e /a) 
\right] 
\delta_{\sigma_{e^-}, \sigma_\nu} \delta( \omega_{e^-}-\omega_\nu-\Delta m)\;.
\label{V2}
\end{eqnarray}
Analogous calculations lead to the following 
amplitudes for  processes $(vi)$ and $(vii)$:
\begin{eqnarray}
{\cal A}^{p  \to n}_{(vi)} & = & 
\frac{4G_F}{\pi a}\sqrt{m_em_\nu \cosh (\pi \omega_{e^+}/a)
\cosh (\pi \omega_{\bar\nu}/a)} \nonumber
\\
& \times & 
Re\left[ 
K_{i\omega_{e^+}/a-1/2}(m_e /a)\; 
K_{i\omega_{\bar\nu}/a +1/2} (m_\nu /a) 
\right] 
\delta_{\sigma_{e^+}, \sigma_{\bar\nu}} 
\delta( \omega_{\bar\nu}-\omega_{e^+}-\Delta m)\;,
\label{VI}
\end{eqnarray}
\begin{eqnarray}
{\cal A}^{p  \to n}_{(vii)} 
& = & 
\frac{4G_F}{\pi a}\sqrt{m_em_\nu \cosh (\pi \omega_{e^-}/a)
\cosh (\pi \omega_{\bar\nu}/a)} \nonumber
\\
& \times & 
Re\left[ 
K_{i\omega_{e^-}/a+1/2}(m_e /a)\; 
K_{i\omega_{\bar\nu}/a +1/2} (m_\nu /a) 
\right] 
\delta_{\sigma_{e^-},- \sigma_{\bar\nu}} 
\delta( \omega_{\bar\nu}+\omega_{e^-}-\Delta m)\;.
\label{VII}
\end{eqnarray}

The differential transition rates   
per absorbed and emitted particle energies 
associated with processes $(v)$-$(vii)$ are given by
\begin{eqnarray}
\frac{1}{T}\frac{d^2 {\cal P}^{p \to n}_{(v)}}{d\omega_{e^-}  d\omega_\nu} 
&=&
\frac{1}{T}\sum_{\sigma_{e^-}=\pm} \sum_{\sigma_{\nu}=\pm}
|{\cal A}^{p  \to n}_{(v)}|^2
n_F(\omega_{e^-}) [1-n_F(\omega_\nu)]\;,
\label{AP1}
\\
\nonumber
\\
\frac{1}{T}
\frac{d^2 {\cal P}^{p \to n}_{(vi)}}{d\omega_{e^+}  d\omega_{\bar\nu}}
&=&
\frac{1}{T}\sum_{\sigma_{e^+}=\pm} \sum_{\sigma_{\bar\nu}=\pm}
|{\cal A}^{p  \to n}_{(vi)}|^2
n_F(\omega_{\bar\nu}) [1-n_F(\omega_{e^+})]\;,
\label{AP2}
\\
\nonumber
\\
\frac{1}{T}
\frac{d^2 {\cal P}^{p \to n}_{(vii)}}{d\omega_{e^-}  d\omega_{\bar\nu}}
&=&
\frac{1}{T}\sum_{\sigma_{e^-}=\pm} \sum_{\sigma_{\bar\nu}=\pm}
|{\cal A}^{p  \to n}_{(vii)}|^2 n_F(\omega_{e^-}) n_F(\omega_{\bar\nu})\;,
\label{AP3}
\end{eqnarray}
where 
\begin{equation}
n_F(\omega) \equiv \frac{1}{1+e^{2\pi \omega/a}}
\label{FT}
\end{equation}
is the fermionic thermal factor associated with the FDU thermal bath
and $T=2\pi\delta (0)$ is the nucleon proper time.
By using Eqs.~(\ref{V2})-(\ref{VII}) in Eqs.~(\ref{AP1})-(\ref{AP3}) 
we obtain 
\begin{eqnarray}
\frac{1}{T} \frac{d^2 {\cal P}^{p \to n}_{(v)}}{d\omega_{e^-}  d\omega_\nu} 
&=&
\frac{4G_F^2}{\pi^3}\left(\frac{m_em_\nu}{a^2}\right) e^{-\pi\Delta m/a} \;
\delta (\omega_{e^-}-\omega_\nu-\Delta m)
\nonumber
\\
&\times&
\left\{Re
\left[ 
K_{i\omega_\nu/a -1/2} (m_\nu/a)
K_{i\omega_{e^-}/a+1/2}(m_e/a)
\right]
\right\}^2\;,
\label{AP4}
\end{eqnarray}
\begin{eqnarray}
\frac{1}{T} 
\frac{d^2 {\cal P}^{p \to n}_{(vi)}}{d\omega_{e^+}  d\omega_{\bar\nu}} 
&=&
\frac{4G_F^2}{\pi^3}\left(\frac{m_em_\nu}{a^2}\right) e^{-\pi\Delta m/a} \;
\delta (\omega_{\bar\nu}-\omega_{e^+}-\Delta m)
\nonumber
\\
&\times&
\left\{
Re\left[ 
K_{i\omega_{\bar\nu}/a +1/2} (m_\nu/a)
K_{i\omega_{e^+}/a-1/2}(m_e/a)
\right]
\right\}^2\;,
\label{AP5}
\end{eqnarray}
\begin{eqnarray}
\frac{1}{T} 
\frac{d^2 {\cal P}^{p \to n}_{(vii)}}{d\omega_{e^-}  d\omega_{\bar\nu}} 
&=&
\frac{4G_F^2}{\pi^3}\left(\frac{m_em_\nu}{a^2}\right) e^{-\pi\Delta m/a} \;
\delta (\omega_{e^-}+\omega_{\bar\nu}-\Delta m)
\nonumber
\\
&\times&
\left\{Re
\left[ 
K_{i\omega_{\bar\nu}/a +1/2} (m_\nu/a)
K_{i\omega_{e^-}/a+1/2}(m_e/a)
\right]
\right\}^2\;.
\label{AP6}
\end{eqnarray}
By integrating Eqs.~(\ref{AP4})-(\ref{AP6}) in frequencies $\omega_{\nu}$
and $\omega_{\bar\nu}$ where it is appropriate, we obtain the following
transition rates associated with each process:
\begin{eqnarray}
\Gamma_{(v)}^{p\to n}
& = & 
\frac{4G_F^2m_e m_\nu }{\pi^3 a^2 e^{\pi\Delta m/a}}  
\int_{\Delta m}^{+\infty} d\omega_{e^-}
\left\{Re\left[ K_{i(\omega_{e^-}-\Delta m)/a -1/2} (m_\nu/a)
                K_{i\omega_{e^-}+1/2}(m_e/a)\right] \right\}^2,
\nonumber
\\
\Gamma_{(vi)}^{p\to n}
& = & 
\frac{4G_F^2m_e m_\nu }{\pi^3 a^2 e^{\pi\Delta m/a}}  
\int_{0}^{+\infty} d\omega_{e^+}
\left\{Re\left[ K_{i(\omega_{e^+}+\Delta m)/a +1/2} (m_\nu/a)
                K_{i\omega_{e^+}-1/2}(m_e/a)\right] \right\}^2,
\nonumber
\\
\Gamma_{(vii)}^{p\to n}
& = & 
\frac{4G_F^2m_e m_\nu }{\pi^3 a^2e^{\pi\Delta m/a}}
\int_{0}^{\Delta m} d\omega_{e^-}
\left\{Re\left[ K_{i(\omega_{e^-}-\Delta m)/a -1/2} (m_\nu/a)
                K_{i\omega_{e^-}+1/2}(m_e/a)\right] \right\}^2.
\nonumber
\end{eqnarray}
We recall that Rindler frequencies may assume 
arbitrary positive real values. (In particular there 
are massive Rindler particles with
zero frequency. See Ref. \cite{HMS} for a discussion on zero-frequency
Rindler particles with finite transverse and angular momentum.) 

The proton decay rate is given by adding up all contributions: 
$
\Gamma_{\rm acc}^{p\to n}= 
 \Gamma_{(v)}^{p\to n} 
+\Gamma_{(vi)}^{p\to n}
+\Gamma_{(vii)}^{p\to n}.
$
This can be written in a compact form as
\begin{equation}
\Gamma_{\rm acc}^{p\to n}= 
\frac{4G_F^2m_e m_\nu }{\pi^3 a^2 e^{\pi\Delta m/a}}  
\int_{-\infty}^{+\infty} d\omega
\left\{Re\left[ K_{i(\omega-\Delta m)/a -1/2} (m_\nu/a)
                K_{i\omega/a+1/2}(m_e/a)\right] \right\}^2.
\label{ATRP}
\end{equation}

At this point we take the limit $m_\nu \to 0$. For this purpose,
it is useful to note that (see (8.407.1), (8.405.1) and  (8.403.1) in 
Ref.~\cite{GR})
\begin{equation}
K_\nu (z) =\frac{\pi}{2\sin \nu \pi}
(J_{-\nu} (iz) e^{i\nu \pi/2} - J_{\nu} (iz) e^{-i\nu \pi/2}) \; ,
\end{equation}
where $\nu $ is noninteger, and $| {\rm arg}\; iz | < \pi $.
Using this expression in conjunction with (8.402) of Ref.~\cite{GR},
we have for small $| z |$ that
\begin{equation}
K_\nu (z) 
\approx 
\frac{\pi}{2\sin \nu \pi}
\left(
(iz/2)^{-\nu}\Gamma^{-1}(-\nu+1) e^{i\nu \pi/2} - 
(iz/2)^{\nu}\Gamma^{-1}(\nu+1) e^{-i\nu \pi/2}
\right) \; .
\label{help}
\end{equation}
By using Eq.~(\ref{help}), we can show that
\begin{equation}
\frac{m_\nu}{a}
K_{i(\omega -\Delta m)/a + 1/2} (m_\nu/a) 
K_{i(\omega -\Delta m)/a - 1/2} (m_\nu/a)
\stackrel{m_\nu \to 0}{\longrightarrow} 
\frac{\pi}{2 \cosh [\pi(\omega - \Delta m)/a]}
\; .
\label{interm2}
\end{equation}
It is now possible to obtain the following 
partial and total transition rates:
\begin{eqnarray}
\Gamma_{(v)}^{p\to n}
& = &
\frac{G_F^2 m_e }{\pi^2 a e^{\pi \Delta m/a}}
\int_{\Delta m}^{+\infty} d\omega_{e^-}
\frac{ K_{i\omega_{e^-}/a + 1/2} (m_e/a)K_{i\omega_{e^-}/a - 1/2} (m_e/a)}
{\cosh[\pi (\omega_{e^-} -\Delta m)/a ] }
 \; ,
\label{R5}
\\
\Gamma_{(vi)}^{p\to n}
& = &
\frac{G_F^2 m_e }{\pi^2 a e^{\pi \Delta m/a}}
\int_{0}^{+\infty} d\omega_{e^+}
\frac{ K_{i\omega_{e^+}/a + 1/2} (m_e/a)K_{i\omega_{e^+}/a - 1/2} (m_e/a)}
{\cosh[\pi (\omega_{e^+} + \Delta m)/a ] }
 \; ,
\label{R6}
\\
\Gamma_{(vii)}^{p\to n}
& = &
\frac{G_F^2 m_e }{\pi^2 a e^{\pi \Delta m/a}}
\int_{0}^{\Delta m} d\omega_{e^-}
\frac{ K_{i\omega_{e^-}/a + 1/2} (m_e/a)K_{i\omega_{e^-}/a - 1/2} (m_e/a)}
{\cosh[\pi (\omega_{e^-} -\Delta m)/a ] }
 \; ,
\label{R7}
\end{eqnarray}
and
\begin{equation}
\Gamma_{\rm acc}^{p\to n}
=
\frac{G_F^2 m_e }{\pi^2 a e^{\pi \Delta m/a}}
\int_{-\infty}^{+\infty} d\omega
\frac{ K_{i\omega/a + 1/2} (m_e/a)K_{i\omega/a - 1/2} (m_e/a)}
{\cosh[\pi (\omega -\Delta m)/a ] }
 \; ,
\label{ATRP2}
\end{equation}
respectively. It is interesting to note that although 
transition rates  have fairly distinct interpretations in the inertial
and accelerated frames, mean proper lifetimes are scalars and must be the same
in both frames. Indeed, by plotting $\tau_p(a)=1/\Gamma_{\rm acc}^{p\to n}$
as a function of acceleration, we do reproduce Fig.~2. In Fig.~3 we plot
the branching ratios 
$$BR_{(v)} \equiv \Gamma_{(v)}^{p\to n}/\Gamma_{\rm acc}^{p\to n},\;\;
BR_{(vi)} \equiv \Gamma_{(vi)}^{p\to n}/\Gamma_{\rm acc}^{p\to n},\;\;
BR_{(vii)} \equiv \Gamma_{(vii)}^{p\to n}/\Gamma_{\rm acc}^{p\to n}.
$$
We note that for small accelerations, where ``few'' high-energy particles
are available in the FDU thermal bath, process $(vii)$ dominates over  
processes $(v)$ and $(vi)$, while for high accelerations, processes $(v)$ and 
$(vi)$ dominate over  process $(vii)$.

A similar analysis can be performed for uniformly accelerated neutrons.  
According to coaccelerated observers, the neutron decay
will be described by a combination of  the following processes:
$$
(viii)\;\; n \; \nu \to p^+ \; e^-\;,\;
(ix) \;\;  n \; e^+ \to p^+ \; \bar \nu\;,\;
(x) \;\; n \to p^+ \; e^- \bar \nu \; .\;\;
$$
The corresponding partial and total transition rates are
\begin{eqnarray}
\Gamma_{(viii)}^{n\to p}
& = &
\frac{G_F^2 m_e }{\pi^2 a e^{-\pi \Delta m/a}}
\int_{\Delta m}^{+\infty} d\omega_{e^-}
\frac{ K_{i\omega_{e^-}/a + 1/2} (m_e/a)K_{i\omega_{e^-}/a - 1/2} (m_e/a)}
{\cosh[\pi (\omega_{e^-} -\Delta m)/a ] }
 \; ,
\label{R8}
\\
\Gamma_{(ix)}^{n\to p}
& = &
\frac{G_F^2 m_e }{\pi^2 a e^{-\pi \Delta m/a}}
\int_{0}^{+\infty} d\omega_{e^+}
\frac{ K_{i\omega_{e^+}/a + 1/2} (m_e/a)K_{i\omega_{e^+}/a - 1/2} (m_e/a)}
{\cosh[\pi (\omega_{e^+} + \Delta m)/a ] }
 \; ,
\label{R9}
\\
\Gamma_{(x)}^{n\to p}
& = &
\frac{G_F^2 m_e }{\pi^2 a e^{-\pi \Delta m/a}}
\int_{0}^{\Delta m} d\omega_{e^-}
\frac{ K_{i\omega_{e^-}/a + 1/2} (m_e/a)K_{i\omega_{e^-}/a - 1/2} (m_e/a)}
{\cosh[\pi (\omega_{e^-} -\Delta m)/a ] }
 \; ,
\label{R10}
\end{eqnarray}
and
\begin{equation}
\Gamma_{\rm acc}^{n\to p}
=
\frac{G_F^2 m_e }{\pi^2 a e^{-\pi \Delta m/a}}
\int_{-\infty}^{+\infty} d\omega
\frac{ K_{i\omega/a + 1/2} (m_e/a)K_{i\omega/a - 1/2} (m_e/a)}
{\cosh[\pi (\omega -\Delta m)/a ] }
 \; ,
\label{ATRN}
\end{equation}
respectively. Fig.~1 gives the neutron mean proper lifetime
$\tau_n(a)=1/\Gamma_{\rm acc}^{n\to p}$ which coincides
with the one calculated in the inertial frame. In Fig.~3
we plot the the branching ratios 
$$BR_{(viii)} \equiv \Gamma_{(viii)}^{n\to p}/\Gamma_{\rm acc}^{n\to p},\;\;
BR_{(ix)} \equiv \Gamma_{(ix)}^{n\to p}/\Gamma_{\rm acc}^{n\to p},\;\;
BR_{(x)} \equiv \Gamma_{(x)}^{n\to p}/\Gamma_{\rm acc}^{n\to p}.
$$
It is easy to see that $BR_{(viii)}=BR_{(v)}$, $BR_{(ix)}=BR_{(vi)}$ and
$BR_{(x)}=BR_{(vii)}$.

\setcounter{equation}{0}


\section{Discussions}
\label{Discussions}

We have analyzed the decay of accelerated protons and neutrons.  
We have compared the particle interpretation of  these decays
in the inertial [see processes $(iii)$ and $(iv)$]
and accelerated [see processes $(v)$-$(vii)$ and $(viii)$-$(x)$] 
frames. They were shown to be quite distinct. Branching ratios
were also evaluated. For protons with small accelerations,
 process $(vii)$ dominates over processes 
$(v)$ and $(vi)$, while for protons with high accelerations 
processes $(v)$ and $(vi)$ dominate over  process $(vii)$.
For neutrons with small accelerations,  
process $(x)$ dominates over processes 
$(viii)$ and $(ix)$, while for neutrons with high accelerations, 
processes $(viii)$ and $(ix)$ dominate over  process $(x)$.
Mean proper lifetimes of the nucleons 
as a function of their proper acceleration
were plotted in Fig.~1 and Fig.~2. For accelerations
such that $a \approx a_c \equiv 2\pi (m_n + m_e - m_p )$ 
we have that $\tau_p \approx \tau_n$.  
Although such accelerations are quite beyond present 
technology, decay of accelerated nucleons might be of some importance
in astrophysics and cosmology. We have performed our calculations
using Fermi theory in a two-dimensional spacetime.
Although this is  suitable  to provide us with  a
qualitative understanding of many conceptual aspects underlying
$\beta$- and inverse $\beta$-decay induced by acceleration, precise 
physical values will only be obtained after a more realistic analysis
is performed. A four-dimensional calculation 
(but restricted to the inertial frame) and its application
to astrophysics and cosmology are presently under
consideration and will be presented somewhere else.

\begin{flushleft}
{\bf{\large Acknowledgements}}
\end{flushleft}
The authors are thankful to Dr. A. Higuchi
for discussions in early stages of this work. 
G.M. was  partially supported by
Conselho Nacional de Desenvolvimento Cient\'\i fico e 
Tecnol\'ogico while D.V. was fully supported by
Funda\c c\~ao de Amparo \`a Pesquisa do Estado de S\~ao Paulo.

\newpage
\begin{figure}
\begin{center}
\mbox{\epsfig{file=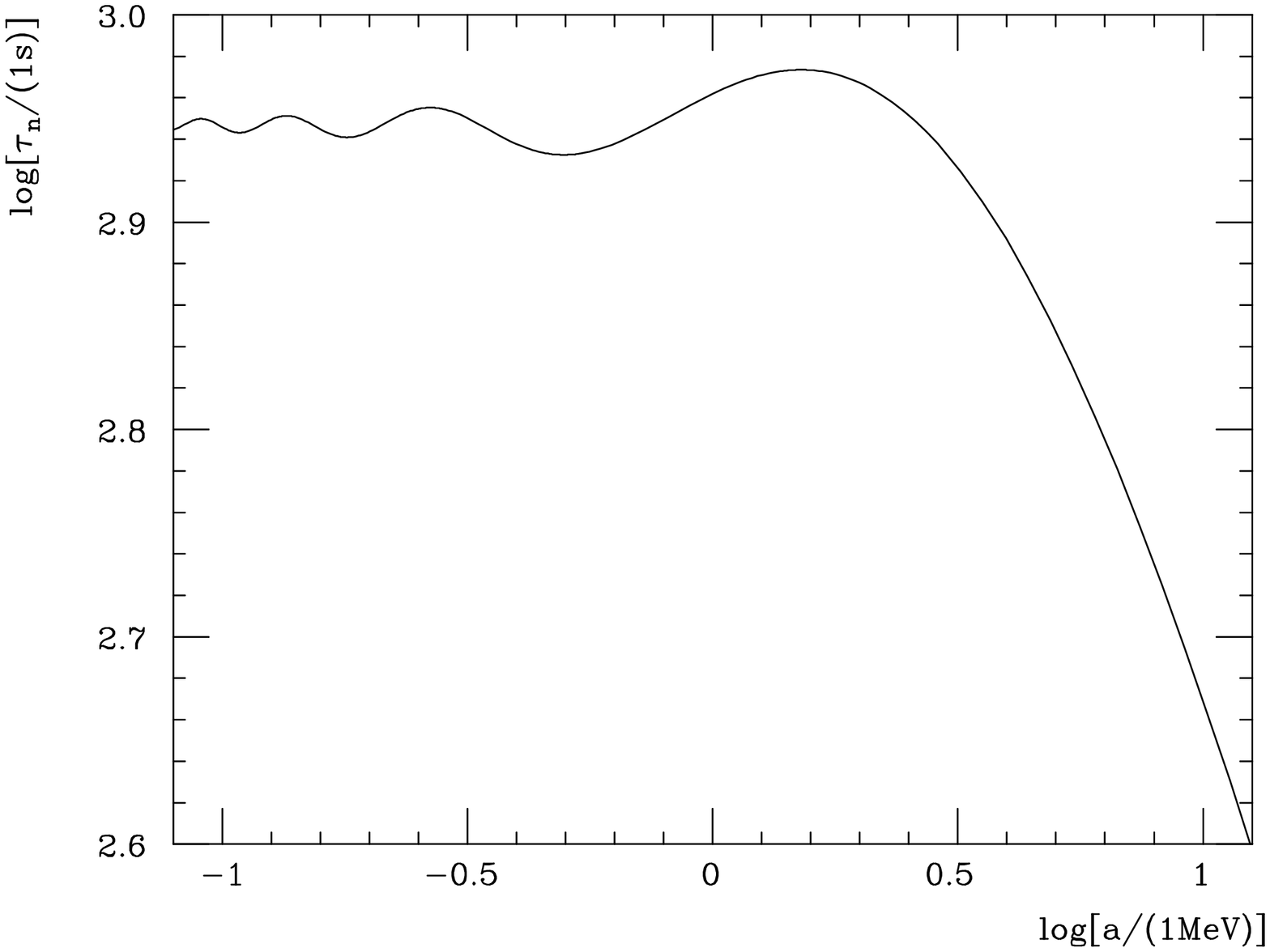,width=0.7\textwidth,angle=90}}
\end{center}
\vskip -1 cm
\caption{ 
The neutron mean proper lifetime is plotted 
as a function of the proper acceleration $a$. Note that
$\tau_n \to 887 s$ as the proper acceleration vanishes.
After an oscillatory regime $\tau_n $ decreases 
steadily as the acceleration increases. 
}
\label{neutron}
\end{figure}

\newpage
\begin{figure}
\begin{center}
\mbox{\epsfig{file=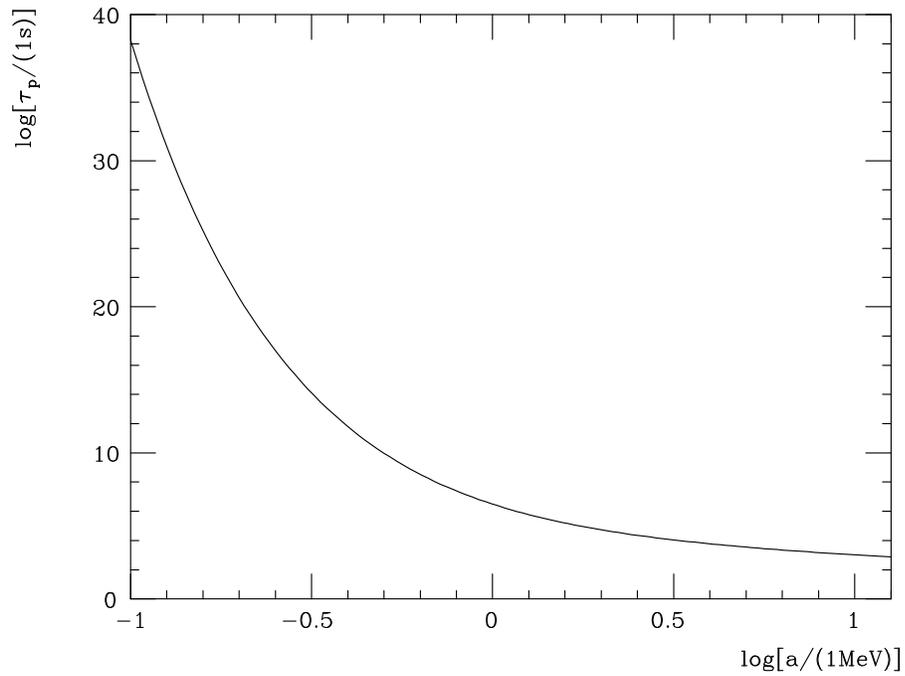,width=0.7\textwidth,angle=90}}
\end{center}
\vskip -1 cm
\caption{ The proton mean proper lifetime is plotted 
as a function of the proper acceleration $a$. 
$\tau_p \to +\infty$ for inertial protons ($a \to 0$). 
For accelerations 
$a  \approx a_c \equiv 2\pi (m_n + m_e - m_p )\approx 11 MeV $ 
we have that $\tau_p \approx \tau_n $. }
\label{proton}
\end{figure}

\newpage
\begin{figure}
\begin{center}
\mbox{\epsfig{file=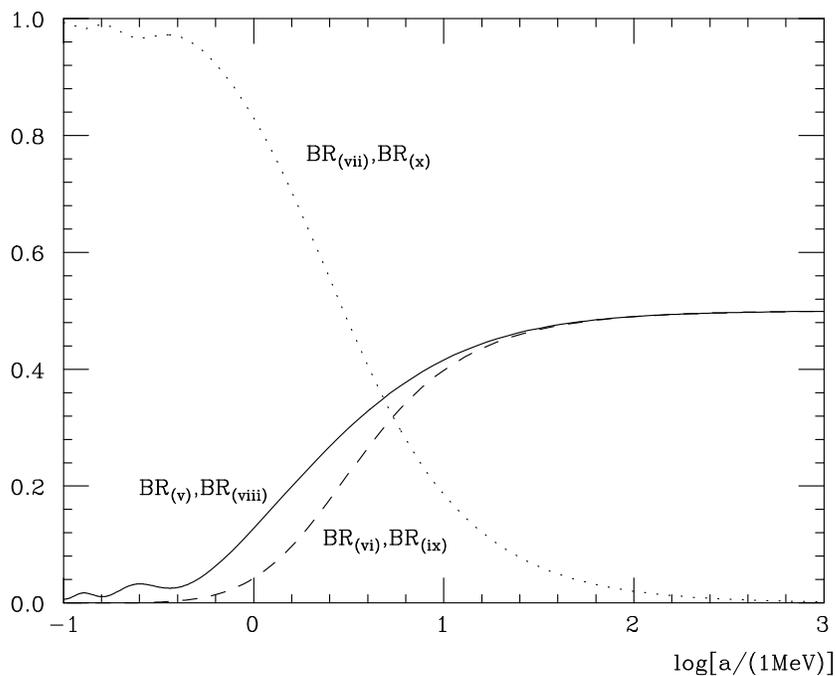,width=0.7\textwidth,angle=90}}
\end{center}
\vskip -1 cm
\caption{Branching ratios 
$BR_{v}, BR_{vi}, BR_{vii}, BR_{viii}, BR_{ix}, BR_{x}$ are plotted. 
For protons, process $(vii)$ dominates over processes 
$(v)$ and $(vi)$ for small accelerations, while  
processes $(v)$ and $(vi)$ dominate over  process $(vii)$
for high accelerations.
For neutrons,  process $(x)$ dominates over processes 
$(viii)$ and $(ix)$ for small accelerations, while  
processes $(viii)$ and $(ix)$ dominate over  process $(x)$ 
for high accelerations.
}
\label{BR}
\end{figure}

\end{document}